# Localised flexural waves in wedges of power-law profile and their relationship with acoustic black holes

Victor V. Krylov[*]

*Department of Aeronautical and Automotive Engineering, Loughborough University, Loughborough, Leicestershire, LE11 3TU, UK*

**ABSTRACT**

In the present paper, the relationship between localised flexural waves in wedges of power-law profile and flexural wave reflection from acoustic black holes is examined. The geometrical acoustics theory of localised flexural waves in wedges of power-law profile is briefly discussed. It is noted that, for wedge profiles with power-law exponents equal or larger than two, the velocities of all localised modes take zero values, unless there is a wedge truncation. It is demonstrated that this effect of zero velocities of localised flexural waves in ideal wedges is closely related to the phenomenon of zero reflection of flexural waves from ideally sharp one-dimensional acoustic black holes. A possible influence of localised wedge modes on flexural wave reflection from one-dimensional acoustic black holes having rough edges is discussed. With regard to two-dimensional acoustic black holes, the role of localised flexural waves propagating along wedge edges that are curved in their middle plane is considered. Such waves can propagate along edges of inner holes in two-dimensional acoustic black holes formed by circular indentations in plates of constant thickness. A possible impact of such localised waves on the processes of scattering of flexural waves by edge imperfections of inner holes in two-dimensional acoustic black holes is discussed, including their influence on the efficiency of two-dimensional acoustic black holes as vibration dampers.

## 1. Introduction

It is well known that solid wedges of power-law profile can support propagation of localised elastic waves along their sharp edges. The existence of such waves in ideal wedges of linear profile (formed by two intersecting planes, see Fig. 1) has been first predicted in 1972 by Lagasse [1] using numerical calculations. Similar numerical predictions have been made independently by Maradudin and co-workers [2], also in 1972, but a few weeks later. These waves, also known as 'wedge elastic waves' or 'wedge acoustic waves', represent a fundamental type of elastic waves in solids, in addition to longitudinal and shear bulk waves and to different types of surface waves, including Rayleigh waves. They contribute to thermal properties of finite solids studied in statistical physics and to modes of vibration of complex wedge-like structures in different engineering applications. They also can be attractive for non-destructive testing of different wedge-like structures.

---

[*]Corresponding author.  *E-mail address*:  V.V.Krylov@lboro.ac.uk

The physical nature of these waves is very complex. According to the more detailed numerical investigations of localised elastic waves in ideal wedges [3] carried out for a material with Poisson's ratio $\sigma = 0.25$, there is a number of antisymmetric modes (for wedges with wedge angles $\theta$ between 0 and 100 degrees) and only one symmetric mode (for wedges with wedge angles $\theta$ between 125 and 180 degrees). The above-mentioned symmetric wedge mode is in fact a modified Rayleigh wave propagating with the velocity that is only slightly lower than the velocity of Rayleigh waves on a flat surface. This mode decays exponentially on both sides from the tip, and it is relatively weakly localised.

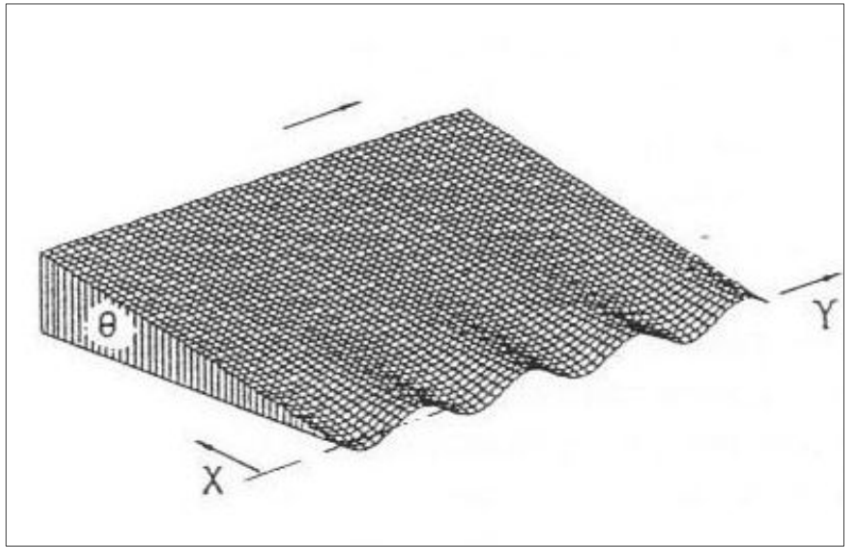

**Fig. 1**. Antisymmetric localised waves propagating along the tip of an ideal linear wedge characterised by the wedge angle $\theta$.

In contrast to the single symmetric mode, the antisymmetric wedge modes are strongly localised, and their velocities can differ very significantly from the velocity of Rayleigh waves, reducing almost to zero for wedges with very small wedge angles $\theta$. For that reason, such antisymmetric modes propagating in slender wedges (characterised by small wedge angles $\theta$) are the most important for practical applications. It should be noted that for wedges of linear profile, that are characterised only by wedge angles $\theta$, the velocities of localised elastic waves do not depend on frequency due to the absence of any characteristic linear dimensions in such wedges. However, localised antisymmetric modes can propagate also in wedges of quadratic profile and in wedges of higher-order power-law profiles. For all such 'nonlinear' profiles, the corresponding localised waves are dispersive, i.e. their phase velocities depend on frequency.

In the present paper, the relationship between localised flexural waves in wedges of power-law profile and flexural wave reflection from acoustic black holes is examined. In particular, it is demonstrated that the effect of zero velocity of localised flexural waves in ideal quadratic wedges is closely related to the phenomenon of zero reflection of flexural waves from ideally sharp one-dimensional acoustic black holes formed by such wedges.

A possible influence of localised wedge modes on flexural wave reflection from one-dimensional acoustic black holes having rough edges is discussed. With regard to two-dimensional acoustic black holes, the existence of localised flexural waves on wedge edges that are curved in their middle plane is considered. Such waves can propagate along edges of the inner holes in two-dimensional acoustic black holes formed by circular indentations in plates of constant thickness. A possible role of such localised waves in the processes of scattering of flexural waves by edge imperfections of inner holes in two-dimensional acoustic black holes is discussed, including their influence on the efficiency of two-dimensional acoustic black holes as dampers of structural vibrations.

## 2. Geometrical acoustics theory of localised elastic waves propagating in slender wedges

In what follows, a brief overview of theoretical and experimental results on wedge elastic waves that are relevant to the main topic of this paper will be presented. First, the geometrical acoustics theory of wedge elastic waves propagating in slender linear wedges (formed by two intersecting planes) will be briefly discussed, based on the results obtained by the present author. Then, this theory will be extended to describe localised wave propagation in wedges of quadratic profile and wedges of more complex geometry. The examples to be considered also include truncated wedges and curved wedges.

*2.1. Localised waves in ideal wedges of linear profile*

Geometrical acoustics approximation is used widely in different problems of classical acoustics for description of sound propagation in ocean and atmosphere [4]. It can be applied also to Lamb waves propagating in plates of variable thickness [5]. The most important modes of Lamb waves are lowest order antisymmetric and symmetric modes, or flexural and compression waves respectively. The latter is often called simply 'plate wave'. In what follows, a particular type of plates of variable thickness will be discussed initially – plates with linearly variable local thickness $h(x)=2\sin(\theta/2) \approx \theta x$ forming slender elastic wedges of linear profile, i.e. wedges with small values of the wedge angle $\theta$ (Fig. 1).

To develop a geometrical acoustics theory of localised wedge waves, one should consider propagation of flexural waves at an arbitrary angle in slender wedges in the geometrical acoustics approximation. To consider this, one can start from the equation for flexural waves in a plate of variable thickness $h(x) \approx \theta x$ [5, 6] (this equation is not shown here for brevity) and seek the solution of this equation for the complex amplitude of the flexural wave $w(x)$ in the following form [5, 6]:

$$w(x) = A(x)e^{i\int [k_a^2(x)-\gamma^2]^{1/2} dx}. \tag{1}$$

Here $k_a(x)$ is the local wavenumber of the propagating flexural wave, $\gamma$ is the projection of the flexural wave vector on the axis $y$ (this projection is considered to be constant), and $A(x)$ represents the wave amplitude that is slowly varying in $x$-direction. It can be shown that equation (1) satisfies the above-mentioned plate equation for flexural waves in a slender wedge asymptotically at relatively high frequencies [5, 6] (see also the monograph [7]) if the function $A(x)$ has the form

$$A(x) = \frac{G}{x\left[k_a^2(x) - \gamma^2\right]^{1/4}}, \quad (2)$$

where $G$ is an arbitrary constant. It can be seen from equation (2) that the amplitude $A(x)$ increases for small values of $x$, when the flexural wave approaches the wedge tip, in agreement with energy conservation law [6]. Relatively high frequencies mean in practice that these frequencies are well within the range of typically used frequencies.

Propagation of localised modes, or guided waves, along the tip of a wedge can be considered as a result of successive reflections of the obliquely incident flexural waves from the tip and their successive turnings back to the tip due to the total internal reflection. The velocities $c$ of the localised wedge modes propagating along the tip of a wedge, i.e. in $y$ direction, can be determined in the geometrical acoustics approximation as solutions of the following Bohr - Sommerfeld type equation [5, 6]:

$$\int_0^{x_t} \left[k_a^2(x) - \gamma^2\right]^{1/2} dx = \pi n, \quad (3)$$

where $\gamma = \omega/c$ is considered as yet unknown wavenumber of a wedge mode, $k_a(x)$ is the already mentioned local wavenumber of a flexural wave in a plate of variable thickness, $n = 1, 2, 3, ...$ is the mode number, and $x_t$ is the so-called ray turning point determined from the equation $k_a^2(x) - \gamma^2 = 0$. In the case of a slender wedge in contact with vacuum the local wavenumber of a flexural wave is $k_a(x) = 12^{1/4} k_p^{1/2} (\theta x)^{-1/2}$, so that $x_t = 2\sqrt{3}k_p/\theta\gamma^2$. Here $k_p = \omega/c_p$ is the wavenumber of a plate compression wave, where $\omega$ is circular frequency, $c_p = 2c_t(1 - c_t^2/c_l^2)^{1/2}$ is the so called plate wave velocity, $c_l$ and $c_t$ are velocities of longitudinal and shear acoustic waves in the plate material.

Taking the integral in equation (3) analytically and solving the resulting algebraic equation yields the extremely simple analytical expression for wedge wave velocities [5-7]:

$$c = \frac{c_p n \theta}{\sqrt{3}}. \quad (4)$$

The expression (4) for velocities of localised wedge modes agrees well with the other (numerically based) theories in the case of slender wedges [1-3] and with the available experimental results. The analytical expressions for amplitude distributions of localised wedge modes are rather cumbersome [5] and are not displayed here. Note that the number of propagating modes of wedge elastic waves depends on the wedge angle $\theta$: the smaller the wedge angle $\theta$, the larger the number of propagating modes $n$. This number can be roughly estimated from the condition $c < c_R$, where $c$ is defined by equation (4) and $c_R$ is Rayleigh wave velocity in the wedge material.

The above-mentioned geometrical acoustics theory of wedge elastic waves can be generalised to consider localised modes in truncated wedges [6], quadratically-shaped elastic wedges [8], wedges immersed in liquids [9], cylindrical and conical wedge-like structures (curved wedges) [10, 11], wedges of general power-law shape [12], wedges made of anisotropic materials [13], and wedges accounting for material nonlinearity [14-16].

*2.2. Localised waves in truncated linear wedges*

This important case relates to real wedges, which are always not ideally sharp but truncated to a certain degree as a result of their manufacturing. This case can be easily analysed using the geometrical acoustics approach to the theory of localised wedge waves described in the previous sub-section. Namely, the velocities $c$ of wedge elastic waves in such structures still can be obtained as a solution of the Bohr - Sommerfeld type equation similar to equation (3):

$$\int_{l}^{x_t} \left[k_a^2(x) - \gamma^2\right]^{1/2} dx = \pi n . \qquad (5)$$

The only difference of equation (5) from equation (3) is that the integration over $x$ is now taken not from zero, but from $l$ indicating the height of the wedge truncation. Performing the integration [6, 7], one can obtain the following algebraic equation in respect of the wedge wave velocities $c$:

$$\frac{c}{c_0}\pi n = \pi n + \left[\frac{2\sqrt{3}k_p l}{\theta}\left(1 - \frac{c_0^2}{c^2}\frac{\sqrt{3}k_p l}{2n^2\theta}\right)\right]^{1/2} + \frac{c}{c_0}n\cos^{-1}\left(1 - \frac{c_0^2}{c^2}\frac{\sqrt{3}k_p l}{n^2\theta}\right), \qquad (6)$$

where $c_0 = c_p n\theta/\sqrt{3}$ are the velocities of wedge modes for an ideal (non-truncated) wedge having the same wedge angle $\theta$, according to equation (4). For ideal wedges (without truncations, i.e. for $k_p l = 0$), equation (6) reduces to equation (4), as expected. In the general case of truncated wedges ($k_p l \neq 0$), the numerical solutions of the equation (6) for the velocities of the first three modes are shown in Fig. 2 by solid curves. For comparison, the corresponding more precise solutions to the thin plate equation for truncated slender wedges earlier obtained using special functions [17] are also shown in Fig. 2 by dashed curves.

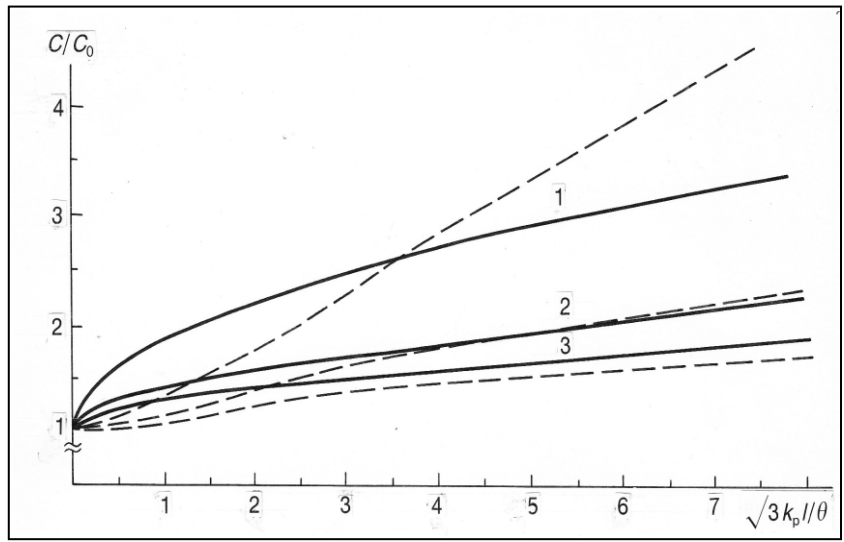

**Fig. 2.** Relative velocities of the first three antisymmetric modes of a truncated linear wedge as functions of the non-dimensional parameter $\sqrt{3}k_p l/\theta$ [6].

It can be seen from Fig. 2 that, according to the solutions obtained using both approaches, all modes of wedge elastic waves in truncated wedges are dispersive, i.e. their velocities depend on frequency. This is what one would expect because, in contrast to the case of ideal wedges, truncated wedges have a characteristic dimension - the length of truncation $l$. One can also see that, for the wedge mode with $n = 1$, the geometrical acoustics solution differs significantly from the more precise solution. This is in agreement with the geometrical acoustics applicability condition for this case [6]. However, for higher-order modes, with $n = 2$ and $n = 3$, the agreement is quite good, especially for large values of the non-dimensional parameter $\sqrt{3}k_p l/\theta$.

*2.3. Localised waves in a linear wedge curved in its own plane*

Using the geometrical acoustics approach, one can consider localised wave propagation in the important case of wedges curved in their own plane [6]. We assume initially that the radius of curvature of the wedge tip is positive (a convex edge) and that it has the value $r_0$ (see Fig. 3).

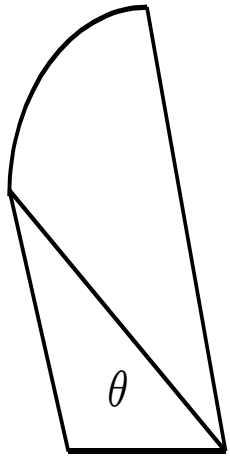

**Fig. 3**. View of a linear convex wedge curved in its own plane (part of a disk of radius $r_0$).

Then, using cylindrical coordinates, in which the edge of a wedge is described by the equation $r = r_0$, one can rewrite the governing Bohr-Sommerfeld type equation (3) in cylindrical coordinates and solve it asymptotically for large $r_0$, i.e. for $r_0 \gg |r_0 - r_t|$. This gives the following approximate expression for the velocities $c$:

$$c = \frac{c_p \theta n}{\sqrt{3}}\left(1 + \frac{\sqrt{3}}{2}\frac{n^2 \theta}{k_p r_0}\right). \tag{7}$$

It can be seen from equation (7) that the velocities of wedge modes $c$ increase in the case of positive curvature (convex edges). If the radius of curvature is negative (concave edges), the velocities $c$ will decrease. In both these cases, the effect of curvature becomes negligible for very large $|r_0|$, when the velocities tend to their values for straight wedges described by equation (4).

It should be noted that equation (7) can be applied also to more complex cases, namely to localised flexural waves propagating along edges of cylindrical and conical wedge-like structures [11]. The results for their velocities for such structures are in good agreement with the earlier performed numerical calculations [10].

## 3. Localised flexural waves in wedges of quadratic profile

An important advantage of geometrical acoustics approach is its ability of analysing localised modes in solid structures of more complex geometry. One of such structures is a truncated wedge of quadratic profile (see Fig. 4). The existence of localised flexural modes in such quadratic wedges has been first predicted by the present author [8]. The local thickness $h(x)$ of a quadratic wedge is described by the function $h(x) = \varepsilon x^2$, where $\varepsilon$ is a dimensional parameter. Using this function in the expression for $k_a(x)$ and substituting the result into the Bohr - Sommerfeld type equation (3), in which the integration should be taken from the value of the truncation length $x_0$ (instead of zero) to the turning point $x_t$, one can derive the following algebraic equation that defines the velocities $c$ of localised modes propagating in a quadratic wedge [8]:

$$\ln\left[z\left(1+\sqrt{1-\frac{1}{z^2}}\right)\right] - \sqrt{1-\frac{1}{z^2}} = \frac{\pi n}{G}. \qquad (8)$$

Here $z = (2\sqrt{3}/\varepsilon x_0 G)(c/c_p)$ and $G = (2\sqrt{3} k_p/\varepsilon)^{1/2}$ are dimensionless variables, $n$ is the mode number, and $k_p = \omega/c_p$. is the wavenumber of a plate compression wave. Note that the condition of applicability of geometrical acoustics approach to this problem is $G >> 1$.

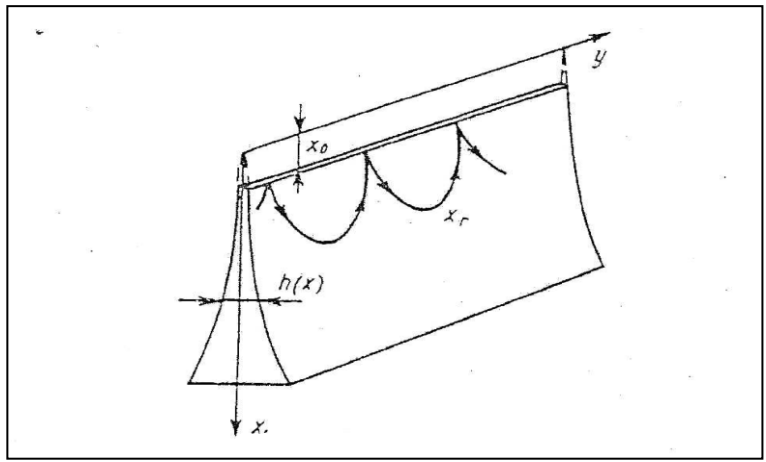

**Fig. 4.** Elastic wedge of quadratic profile: $h(x) = \varepsilon x^2$; solid lines with arrows illustrate propagation of localised flexural waves in y-direction as a series of successive reflections from the tip and turning back due to the total internal reflection.

The equation (8) is rather complicated, and in general case it can be solved only numerically. However, its approximate analytical solutions can be also obtained for some limiting conditions. In particular, for $z >> 1$ the approximate solution for the normalised velocities of localised wedge modes takes the form [8]

$$\frac{c}{c_p} = \frac{\varepsilon x_0}{4\sqrt{3}} G \exp\left(1 + \frac{\pi n}{G}\right). \tag{9}$$

It can be seen from equation (9), that with the increase in $G$ the ratio $c/c_p$ first decreases to a certain minimum and then increases linearly with $G$. It is important to note that the values of $c/c_p$ in equation (9) are proportional to the dimensionless parameter $\varepsilon x_0$. It follows from this proportionality that the values of $c/c_p$ tend to zero as $x_0 \rightarrow 0$, i.e. in the case of ideal (non-truncated) wedges of quadratic profile.

Typical results of the numerical solution (by the bisection method) of the equation (8) for relative velocities of localised modes $c/c_p$ as functions of the non-dimensional parameter $G = (2\sqrt{3}k_p/\varepsilon)^{1/2}$, where $k_p = \omega/c_p$, are shown in Fig. 5. Calculations have been carried out for the value of the non-dimensional parameter $\varepsilon x_0/2\sqrt{3}$ (involving the truncation length $x_0$) equal to 0.005. It can be seen from Fig. 5 that wedge elastic waves in quadratic wedges are dispersive, which reflects the fact that truncated quadratic wedges are characterised by the two dimensional parameters: $\varepsilon$ and $x_0$.

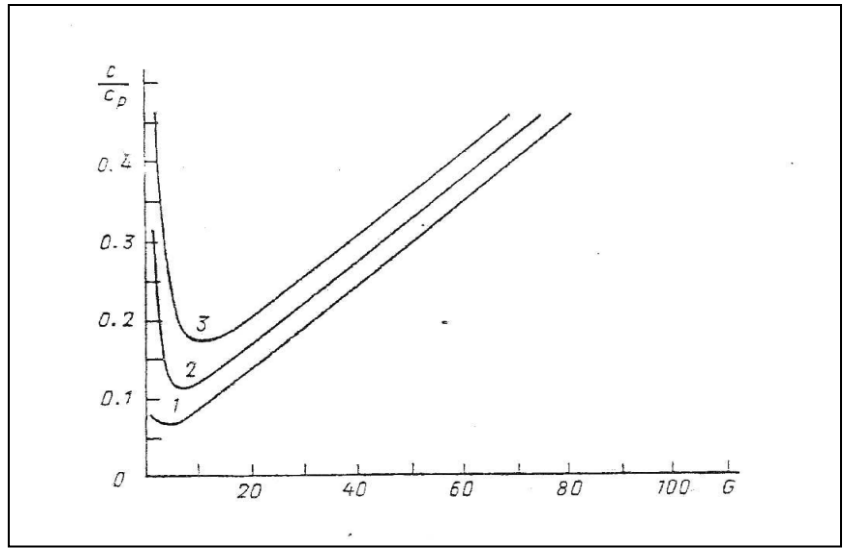

**Fig. 5**. Dispersion curves of the three lowest order modes of a quadratic wedge as functions of the non-dimensional parameter $G = (2\sqrt{3}\omega/\varepsilon c_p)^{1/2}$ [8].

It can be seen from Fig. 5 that, in agreement with the above-mentioned approximate analytical solution, the mode velocities initially decrease with frequency, which is proportional to $G^2$, until they reach their minimum values, before starting to increase. Note that the phase velocities of all these modes are proportional to the dimensionless parameter $\varepsilon x_0$ describing the wedge truncation. Obviously, by reducing the value of the parameter $\varepsilon x_0$, e.g. by reducing the length of truncation $x_0$, the velocities of antisymmetric (flexural) wedge modes can be made arbitrary small. In particular, for $x_0 = 0$ corresponding to the case of ideal quadratic wedge (without truncation), the velocities of all modes are zero.

It can be shown that velocities of localised flexural modes become zero not only in quadratic wedges, but also in wedges of higher-order power-law profiles, i.e. if the exponent $m$ in the function of wedge local thickness $h(x) = \varepsilon x^m$ is equal or larger than two [12].

## 4. Acoustic black holes and localised flexural waves in elastic wedges

### 4.1. The physical principle of acoustic black holes

As it is known, the physical principle of black holes for waves of any physical nature can be explained by considering a general case of one-dimensional wave propagation characterised by the distance $x$ in an ideal medium with power-law dependence of wave velocity $c$ on $x$ as $c = ax^n$, where $n$ is a positive rational number and $a$ is a constant [18]. One can express the geometrical acoustics solution for the complex amplitude $U(x)$ of a wave propagating from any arbitrary point $x$ towards a zero point (where $c = 0$) as

$$U(x) = A(x)e^{i\Phi(x)}. \tag{10}$$

Here

$$\Phi = -\int_x^0 k(x)dx = \int_0^x k(x)dx \tag{11}$$

is the total accumulated phase, and $A(x)$ is a slowly varying amplitude. Since $k(x) = \omega/c(x) = \omega/ax^n$, it is obvious from equation (11) that for $n \geq 1$ the integrals in (11) diverge and the phase $\Phi$ becomes infinite. This means that under these circumstances the wave never reaches the edge. Therefore, it never reflects back either, i.e. the wave becomes trapped, thus indicating that the above mentioned ideal medium with a linear or higher power-law profile of wave velocity can be considered as 'acoustic black hole' for the wave under consideration.

This phenomenon of zero reflection has been first described by Pekeris [19] for underwater sound waves in a stratified ocean with sound velocity profile linearly decreasing to zero. Mironov [20] was the first to predict a practically important possibility of zero reflection of flexural waves from a tip of an ideal quadratic wedge. Note in this connection that a quadratic wedge provides the required minimum linear decrease in flexural wave velocity towards a sharp edge.

Let us now consider the simplest one-dimensional case of plane flexural wave propagation in the normal direction towards the edge of a free elastic wedge described by a power-law relationship $h(x) = \varepsilon x^m$, where $m$ is a positive rational number and $\varepsilon$ is a constant (Fig. 6). Since flexural wave propagation in such wedges can be described in the geometrical acoustics approximation [5], the total accumulated phase $\Phi$ resulting from the wave propagation from an arbitrary point $x$ located in the wedge medium plane to the wedge tip ($x = 0$) can be expressed by the above-mentioned equation (11). In this case $k(x)$ is a local wavenumber of a flexural wave for a wedge in contact with vacuum: $k(x) = 12^{1/4} k_p^{1/2} (\varepsilon x^m)^{-1/2}$, where $k_p = \omega/c_p$ is the wavenumber of a plate compression wave, $c_p = 2c_t(1-c_t^2/c_l^2)^{1/2}$ is its phase velocity, $c_l$ and $c_t$ are longitudinal and shear wave velocities in a wedge material, and $\omega = 2\pi f$ is circular frequency. Again, one can easily see that the integral in equation (11) diverges for $m \geq 2$. This means that the phase $\Phi$ becomes infinite and the wave never reaches the edge.

Therefore, it never reflects back either, thus indicating that the above mentioned ideal wedges represent acoustic 'black holes' for incident flexural waves.

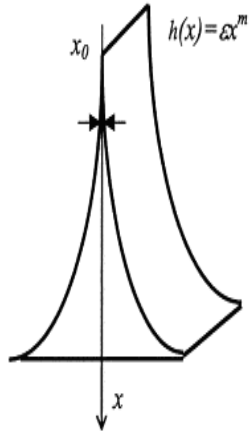

**Fig. 6**. Geometry of an elastic wedge of power-law profile; a possible truncation is characterised by the truncation length $x_0$.

As was already mentioned, in contrast to the ideal wedge of power-law profile that was described above, real fabricated wedges always have truncated edges, and this adversely affects their performance as acoustic black holes, which also depends on material attenuation. For typical (rather low) values of material attenuation in such materials as steel, even very small truncations result in reflection coefficients $R_0$ being as large as 50-70 % [20], which makes it impossible to use such wedges as practical vibration dampers. It has been found though [21-23] that the performance of real wedges (with truncations) can be drastically improved via increasing wave energy dissipation in the area of slow wave velocity (near the sharp edges) by covering wedge surfaces near the edges by thin absorbing layers (films), e.g. by polymeric films, which can reduce the reflection coefficients down to 1-3 %. Such improved black hole structures have promising applications for vibration damping, and they continue to be widely investigated (see e.g. [24-27]).

In addition to the above-mentioned wedges of power-law profile, which can be considered as one-dimensional acoustic black holes attached to the edges of plates or beams, two-dimensional acoustic black holes have been also proposed [28]. Two-dimensional acoustic black holes are protruding cylindrically symmetrical indentations (pits) of power-law profile (Fig.7) drilled in a regular thin plate of constant thickness. The central area of the pits is usually covered by a small piece of absorbing layer to make them more efficient vibration absorbers taking away part of the energy of the incident flexural wave. Such two-dimensional acoustic black holes and their combinations are also subject to extensive theoretical and experimental research (see e.g. [18, 29-34]).

The above-mentioned two-dimensional acoustic black holes (power-law pits) can be placed at any point of a plate or any other plate-like or shell-like structure. The effect of such black hole is in eliminating some flexural wave rays, intersecting with the black hole, from contributing to the overall frequency response function of a structure, which may result in substantial damping of some resonant peaks in the frequency response function. To amplify the vibration damping effect of two-dimensional acoustic black holes one can place ensembles of several black holes distributed over the structure (e.g. periodic arrays of black holes).

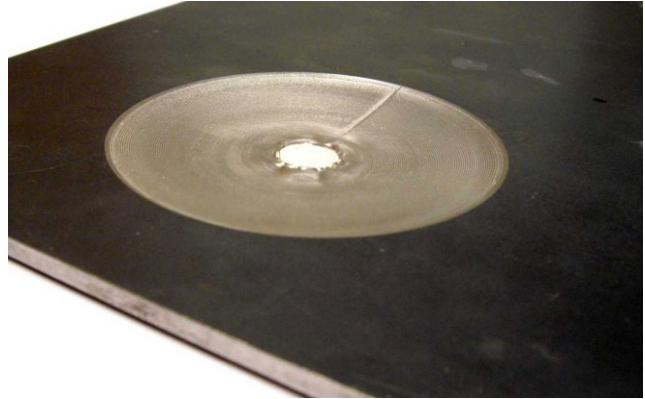

**Fig. 7**. View of a manufactured two-dimensional acoustic black hole embedded into a rectangular steel plate (absorbing material is not shown); one can clearly see imperfections on the edge of the inner hole.

One of the most important advantages of the above-mentioned one-dimensional and two-dimensional acoustic black holes as dampers of structural vibrations is that they are efficient even for relatively thin and small pieces of attached absorbing layers. The reason for this is that wave energy dissipation takes place mainly in a very small area near sharp edges.

*4.2. Comparison with localised wave propagation along the tips of elastic wedges*

Obviously, the above-mentioned effect of zero reflection of flexural waves is linked to the predicted zero values of localized wave velocities in quadratic and higher order elastic wedges [8, 12]. This can be seen from the comparison of equations (10) and (11) with equation (1). The latter represents a combination of equations (10) and (11) and differs from them only in the fact that it describes the propagation of flexural waves at oblique angle in respect of y-axis, which is described by the non-zero projection $\gamma$ of the flexural wave vector on the axis $y$ (this projection is considered to be constant).

When an incident flexural wave described by equation (1) approaches the tip of an ideal (non-truncated) wedge the absolute value of $k_a(x)$ tends to infinity, and so does the x-projection of the flexural wave vector, so that the wave approaches the edge effectively at a normal angle, like in the case of one-dimensional problem described by equations (10) and (11). Thus, the above-mentioned effect of zero reflection at normal incidence of flexural waves also takes place in the case of their oblique incidence. This means that, considering the propagation of localised flexural waves along the tip of an ideal quadratic wedge as a series of successive reflections from the tip and successive processes of total internal reflection (see Fig. 4), one can conclude that already in the process of the first reflection from the wedge tip it will take an infinite time for the wave to reach the tip, which means that the wave will never reach it and thus will never reflect back, which will result in the absence of successive reflections. All these can be interpreted as propagation of a localised wave along the wedge tip with zero velocity.

In the case of truncated quadratic wedges having very small values of truncation length $x_0$ and having no material losses, there will be successive reflections (with the amplitudes of the reflection coefficient being equal to unity), but the reflected waves will be substantially delayed because the values of the total accumulated phase for each reflection will remain very large. This will result in propagation of localised flexural waves with non-zero velocities, which can be arbitrary small, depending on the value of $x_0$. Thus, the effect of zero velocities of localised flexural waves propagating along the tips of ideal (non-truncated) wedges of quadratic and higher-order profiles is linked to the effect of zero reflection of flexural waves from the tips of the same wedges considered in this case as acoustic black holes. In this connection, one can say that the effect of zero velocities of localised flexural waves propagating in elastic wedges of power-law profile is another manifestation of the acoustic black hole effect, which is usually associated with zero reflection.

## 5. Possible contributions of localised wedge modes to acoustic black hole performance

In addition to the above-mentioned fundamental relationship between zero velocities of localised elastic waves propagating in wedges of power-law profile and zero reflection of flexural waves from acoustic black holes, a question arises about possible contributions of localised wedge modes to the performance of real one-dimensional and two-dimensional acoustic black holes.

Let us first consider one-dimensional acoustic black holes (elastic wedges of power-law profile). It follows from the Snell's law that localised wedge modes, which should have very low propagation velocity in wedges in the case of small truncations, cannot be generated in the process of reflection of flexural waves from the tips of such wedges. However, real wedges are not only truncated, but they also have numerous imperfections on the tips as a result of their manufacturing. Scattering of the incident flexural waves on such imperfections can result in generation of localised wedge modes and in their subsequent multiple scattering into all types of elastic waves, including flexural waves, which may result in additional energy losses of the incident flexural waves resulting in improving the performance of acoustic black holes as vibration dampers.

The effects of deviations of manufactured wedge-like structures from ideal elastic wedges of power-law profile on damping flexural vibrations have been investigated in the papers [35-37]. In particular, the effect of mechanical damage to wedge tips has been studied, including tip roughness, tip curling and tip extension. It has been found that such deviations are not detrimental for damping performance. On the contrary, in many cases they improve the performance due to additional energy losses. Although localised wedge modes have not been explicitly mentioned in these investigations, it is quite likely that they play their part in improving vibration damping performance of such acoustic black holes. A more detailed investigation would be required to achieve better understanding of these processes.

Similarly, the case of two-dimensional black holes (circular indentations of power-law profile) can be considered. The presence of imperfections on the edge of the inner hole in a real manufactured two-dimensional acoustic black hole indentation can be clearly seen in Fig. 7. The edge of this internal hole can be considered as a concave wedge of power-law profile curved in its own plane. Like in the case of linear curved wedges that have been discussed in Section 2.3, such power-law wedges formed by the edges of inner holes can support propagation of localised wedge modes that can be excited via scattering of incident flexural waves on edge imperfections. Again, one could expect that additional energy losses caused by this scattering would improve vibration damping performance of two-dimensional acoustic black holes.

## 6. Conclusions

In the present paper, the geometrical acoustics theory of localised flexural waves propagating along the tips of elastic wedges of power-law profile has been briefly discussed. The emphasis has been made to localised modes of linear wedges (formed by two intersecting planes) and wedges of quadratic profile. In particular, it has been demonstrated that the velocities of all localised modes in wedges of quadratic and higher-order profiles are zero, unless there is a truncation on the wedge tip. In the latter case, the velocities can be arbitrary small, depending on the value of the truncation length.

It has been shown that the effect of zero velocities of localised flexural waves propagating along the tips of ideal (non-truncated) wedges of quadratic and higher-order profiles is linked to the effect of zero reflection of flexural waves from the tips of the same wedges considered in this case as acoustic black holes. In this connection, one can say that the effect of zero velocities of localised flexural waves propagating in elastic wedges of power-law profile is another manifestation of the acoustic black hole effect, which is usually associated with zero reflection.

Possible contributions of localised wedge modes to the performance of one-dimensional and two-dimensional acoustic black holes have been briefly discussed. It is expected that such localised modes can be generated via scattering of incident flexural waves on the imperfections of wedge tips (in one dimensional acoustic black holes) or edges of inner holes (in two-dimensional acoustic black holes), which may result in additional energy losses resulting in improving the vibration damping performance of such black holes. A more detailed investigation would be required to achieve better understanding of these processes.